\documentclass[aps,prl,amsfonts,nofootinbib,superscriptaddress,twocolumn]{revtex4}
\usepackage{amsmath}
\usepackage{graphicx}

\newcommand{\beq}{\begin{equation}}
\newcommand{\eeq}{\end{equation}}
\newcommand{\bea}{\begin{eqnarray}}
\newcommand{\eea}{\end{eqnarray}}
\newcommand{\dk}{{\int \frac{\d^3 \vc{k}}{(2\pi)^{3/2}}}}

\newcommand{\vc}[1]{{\textbf{\em #1}}}

\newcommand{\pmtrx}[1]{\begin{pmatrix} #1 \end{pmatrix}}

\newcommand{\der}{\partial}
\renewcommand{\d}{\mathrm{d}}
\newcommand{\non}{\nonumber}
\DeclareMathSymbol{\mg}{\mathrel}{symbols}{"1D}

\newcommand{\inv}{^{-1}}

\newcommand{\ga}{\alpha}
\newcommand{\gb}{\beta}

\newcommand{\gd}{\delta}
\renewcommand{\ge}{\epsilon}

\newcommand{\gz}{\zeta}
\newcommand{\get}{\eta}
\newcommand{\gth}{\theta}

\newcommand{\gk}{\kappa}

\newcommand{\gx}{\xi}

\newcommand{\gr}{\rho}

\newcommand{\gf}{\phi}

\newcommand{\gc}{\chi}

\newcommand{\gD}{\Delta}

\newcommand{\gP}{\varPi}

\newcommand{\cD}{{\mathcal D}}

\newcommand{\cW}{{\mathcal{W}}}

\newcommand{\ta}{{\tilde a}}

\newcommand{\tf}{{\tilde f}}

\newcommand{\tV}{{\tilde V}}

\newcommand{\tga}{{\tilde\alpha}}

\newcommand{\tgz}{{\tilde\zeta}}

\newcommand{\bA}{{\bar A}}

\begin{document}

\title{Quantitative bispectra from multifield inflation}

\author{G.I.~Rigopoulos}
\affiliation{Institute for Theoretical Physics, Utrecht University,
Postbus 80.195, 3508 TD Utrecht, Netherlands}
\author{E.P.S.~Shellard}
\affiliation{Department of Applied Mathematics and Theoretical Physics,
University of Cambridge,\\
Centre for Mathematical Sciences, Wilberforce Road, Cambridge CB3 0WA,
United Kingdom}
\author{B.J.W.~van Tent}
\affiliation{Laboratoire de Physique Th\'eorique, Universit\'e Paris-Sud 11,
B\^atiment 210, 91405 Orsay, France}

\begin{abstract}

\noindent After simplifying and improving the non-Gaussian formalism we
developed in previous work, we derive a quantitative expression for
the three-point correlator (bispectrum) of the curvature perturbation in 
general multiple-field inflation models. Our result describes the evolution
of non-Gaussianity on superhorizon scales caused by the nonlinear
influence of isocurvature perturbations on the adiabatic
perturbation during inflation. We then study a simple quadratic
two-field potential and find that when slow roll breaks down and the
field trajectory changes direction in field space, the
non-Gaussianity can become large. However, for the simple models studied
to date, the magnitude of this non-Gaussianity decays away after the
isocurvature mode is converted into the adiabatic mode.

\end{abstract}

\maketitle

\section{Introduction}
The comparison of inflationary predictions with increasingly precise
observations has
become a major focus of cosmology over the past decade.
Observations of the cosmic
microwave background (CMB), notably by WMAP and the forthcoming
Planck satellite, provide accurate measurements of the CMB
temperature anisotropies. If this experimental effort is to place
any further constraints on inflation, it should be matched by a
corresponding effort to improve the accuracy of theoretical predictions.
To date, the study of linearized perturbations around a homogeneous
background has provided an excellent first approximation. However,
gravity is inherently nonlinear so it is important to investigate
whether higher-order effects are relevant for future high-precision
observations.

Any nonlinearity will manifest itself as non-Gaussianity in the
correlations or other measures of the stochastic properties of the
CMB anisotropies, a prime measure being the three-point correlator or
bispectrum. The general expectation has been that non-Gaussianity
generated during inflation will be very small and unobservable. In
this paper, summarizing and extending results from our two earlier papers
\cite{mf, formalism},
we quantitatively study this question for multifield
inflation and apply our results to models with two
scalar fields, evaluating our expressions numerically. We find that
non-Gaussianity evolves due to the influence of isocurvature
perturbations on the adiabatic perturbation and can become large
when slow roll breaks down and the field trajectory changes
direction in field space --- this happens when the heavier field rolls
down to reach its minimum value and the lighter field takes over. However,
in the simple cases studied to date, this large
non-Gaussianity is temporary and decays after the turn of the field
trajectory, as the conversion of the isocurvature modes is completed.

The effort to characterize
non-Gaussianity from inflation has attracted a great deal of
interest  because it may open up a new observational window on the
early universe (see e.g.\ the review \cite{review} and more recent
papers \cite{ng-lit,wandsvernizzi}). This effort has been successful in
investigating non-Gaussianity in a variety of specific regimes, but
in this paper we focus on the important influence of the
isocurvature modes on the adiabatic mode during inflation.
First, we introduce the basic formalism using our key gradient variable
$\zeta_i$ to characterise nonlinear inhomogeneities in the long-wavelength
approximation.  Next, we identify a constraint between the first isocurvature
mode and the adiabatic time derivative, which allows us to reduce the
$2n$ nonlinear evolution equations to an elegant $2n$$-$$1$ system (with
$n$ fields).  In the two-field case, we then perturbatively expand these 
equations to second order and we obtain an integral expression for the 
three-point correlator in terms of linear solutions and background slow-roll 
parameters (at no stage requiring these to be small).  A numerical 
investigation of a two-field quadratic model then demonstrates how a 
potentially large bispectrum emerges and decays.

\section{Nonlinear inhomogeneous quantities}

We begin by summarising and extending the formalism developed in
\cite{formalism} to compute the nonlinear evolution on superhorizon
scales during multifield inflation. Spacetime is described by the
long-wavelength metric
    \beq
    \d s^2=-N^2(t,\vc{x})\d t^2+a^2(t,\vc{x})\d\vc{x}^2,
    \eeq
with the lapse function $N(t,\vc{x})$, defining the time slicing,
and the scale factor $a(t,\vc{x})$ both dependent on space and
time. The key variable describing the full nonlinear
inhomogeneities is \cite{formalism}
    \beq\label{gi_var}
    \gz_i^m \equiv \gd_{m1} \der_i \ln a
    - \frac{\gk}{\sqrt{2\ge}} \, (e_{mA}\, \der_i \gf^A)\,,\
    \eeq
with $\gk^2 \equiv 8\pi G = 8\pi/m_\mathrm{pl}^2$ and $\gf^A$ the
scalar fields of the multifield inflation model; $\ge$ and $e_m^A$
are defined below. Note that we allow for the possibility of a
nontrivial field metric $G_{AB}$, used to raise and lower indices
in field space. The Einstein summation convention is assumed
implicitly throughout this paper. The nonlinear quantity $\gz_i^m$
transforms as a scalar under changes of time slicing on long
wavelengths \cite{formalism}; under the transformation
 $X^\mu=(t,\vc{x})\rightarrow
\tilde{X}^\mu=(T(t,\vc{x}),\vc{x})$,
    \beq\label{transform}
    \gz^m_i(t,\vc{x})=\tgz^m_i(T,\vc{x}).
    \eeq
Here, the spatial part
of the transformation is independent of time and can be taken
to be trivial \cite{formalism}.
In the single-field case and when linearised, (\ref{gi_var}) is just
the spatial gradient of the well-known curvature perturbation $\gz$
in the literature.  Note that our variable (\ref{gi_var}) has been
further studied in a covariant framework in \cite{LV} and its
relationship to perturbation theory elaborated.

We have chosen an orthonormal set of basis vectors $e_m^A$ in field
space, which are defined from the field velocity, $e_1^A \equiv
\gP^A/\gP$, and successively higher-order time derivatives via an
orthogonalization process \cite{vantent1,vantent}. Here
$\gP^A\equiv\dot{\gf}^A/N$ is the proper field velocity, with
length~$\gP$. For example, $e_2^A = (\delta^A_B-e_1^Ae_{1B}) \cD_t
\gP^B/|\ldots|$, with $\cD_t$ the time derivative covariant with
respect to the metric $G_{AB}$ in field space. A similar
orthogonalization procedure for the special case of two fields and a
flat field metric was independently proposed in \cite{gordon}. This
basis has the advantages of immediately separating the adiabatic
component $\gz_i^1$ from the isocurvature modes $\gz_i^m\; (m\geq
2)$ and making the origin of non-Gaussianity more transparent. We
also define the {\it local} slow-roll parameters
\cite{vantent1,vantent}
    \bea\label{slowrollpar}
    \ge(t,\vc{x}) & \equiv & \frac{\gk^2 \gP^2}{2 H^2},
    \qquad
    \get^\parallel(t,\vc{x}) \equiv - 3 - \frac{\gP^A  V_{,A}}{H\gP^2},
    \non\\
    \get^\perp(t,\vc{x}) & \equiv & - \frac{e_2^A V_{,A}}{H\gP},
    \quad
    \gc(t,\vc{x}) \equiv \frac{V_{22}}{3 H^2}
    + \ge + \get^\parallel,
    \\
    \gx_m(t,\vc{x}) & \equiv & - \frac{V_{m1}}{H^2}
    + 3(\ge-\get^\parallel)\gd_{m1} - 3\get^\perp \gd_{m2},
    \non
    \eea
where $V(\gf^A)$ is the potential of the model, with its covariant
field derivatives $V_{,A} \equiv  \der_A V$ and $V_{;AB} \equiv
\cD_B\der_A V$. Furthermore, $V_{mn}\equiv e_m^A e_n^B V_{;AB}$ and
$H\equiv\dot{a}/(Na)$ is the Hubble parameter. We have chosen the number of
e-folds $t=\ln a$ (i.e.\ $NH=1$) as the time variable, which simplifies the
equations. We emphasize that we are \emph{not} making any slow-roll
approximations, the slow-roll parameters are just short-hand notation and can
be large.

In the definitions~(\ref{slowrollpar}), $\ge$ and
$\get^\parallel$ are the effectively single-field slow-roll
parameters. The truly multiple-field slow-roll parameter $\get^\perp$
is proportional to the component of the field acceleration perpendicular to
the field velocity; i.e.\ it is non-zero only when the trajectory in
field space makes a turn (at least for a flat field manifold;
if the manifold is curved even a straight trajectory can have
$\get^\perp\neq 0$ due to the connection terms). 
To physically interpret $\chi$ we rewrite it as
$3H^2 \chi = V_{22} - V_{11} + 6H^2 \ge$, valid to leading order in slow roll.
Thus, $\chi$ is related to the difference between the effective mass
perpendicular to the field trajectory and the one parallel to it, with a
correction proportional to $H^2 \ge$, which is actually dominant
initially. In this language $3H^2 \get^\perp = -V_{21}$ (leading-order
slow roll) is proportional to the off-diagonal mass~$V_{21}$. The $\gx_m$
are second-order slow-roll parameters.

In \cite{formalism}, the field and Einstein equations were
reformulated as an equation of motion for the inhomogeneous
variable $\gz_i^m$. We combine $\gz_i^m$ and
its time derivative $\gth_i^m \equiv \der_t(\gz_i^m)$ into a
single vector,
   \beq
   w_{i\,\ga}\equiv(\gz_i^1,\gth_i^1,\gz_i^2,\gth_i^2,
   \gz_i^3,\gth_i^3,\ldots)^T\,,
   \eeq
with $T$ denoting the transpose. (The reason for the change of notation with
respect to \cite{mf} will become clear below.) Then the nonlinear evolution
equation, valid on superhorizon scales, becomes
    \beq\label{basic_sym4}
    \dot{w}_{i\,\ga}(t,\vc{x}) + B_{\ga\gb}(t,\vc{x}) w_{i\,\gb}(t,\vc{x})
    = 0,
    \eeq
where the index $\ga$ runs over twice the number of fields. This
equation is exact within the long-wavelength approximation where
second-order spatial gradients are neglected and is valid beyond
slow roll. The full expression for the matrix $B_{\ga\gb}$ can be
found in \cite{mf}; for two fields on a flat manifold in the
$NH=1$ gauge it is exactly
    \beq
    B = \pmtrx{0&-1&0&0\\0&3+\ge+2\get^\parallel
    &-6\get^\perp-4\ge\get^\perp-2\gx^\perp&-2\get^\perp\\
    0&0&0&-1\\0&2\get^\perp
    &3\gc+2\ge^2+4\ge\get^\parallel+\gx^\parallel
    &3+\ge+2\get^\parallel},
    \eeq
where $\gx^\parallel \equiv \gx_1$ and $\gx^\perp \equiv \gx_2$.
Since the matrix $B$ is spatially dependent via $H(t,\vc{x})$,
$\gf^A(t,\vc{x})$, and $\gP^A(t,\vc{x})$ [see
(\ref{slowrollpar})], the nonlinear system (\ref{basic_sym4}) has
to be closed with relations expressing these quantities in terms
of $w_{i\,\ga}$. Again, the exact constraint equations, derived from
the Einstein equations, can be found in \cite{mf}, but in the
two-field case they become
    \bea\label{constr}
    \der_i \ln H & = & \ge \, \gz_i^1,
    \qquad
    e_{m\,A} \der_i \gf^A \: = \: - \frac{\sqrt{2\ge}}{\gk} \, \gz_i^m,
    \\
    e^A_m \cD_i \gP_A & = & - \frac{H\sqrt{2\ge}}{\gk}
    \Bigl( \gth_i^m + \get^\parallel \gz_i^m - \get^\perp \gz_i^2 \gd_{m1}
    \non\\
    && \qquad\qquad + (\get^\perp \gz_i^1 + \ge \gz_i^2)\gd_{m2} \Bigr).
    \non
    \eea

It turns out that there is an exact algebraic relation between
$\gth_i^1$ and $\gz_i^2$, which allows us to reduce the order of the
system by one. The derivation of this relation goes as follows.
Taking the $m=1$ component of the $\cD_i \gP^A$ constraint and
multiplying by $-\gk/(H\sqrt{2\ge})$, we find that the left-hand
side equals $-\der_i \ln \gP$. Taking the time derivative of this
equation and comparing with the equation of motion for $\gth_i^1$
[the $\ga=2$ component of (\ref{basic_sym4})], using the fact that
$\der_t \ln \gP = \get^\parallel$ as can be seen from the field
equation, $\cD_t \gP^A = -3\gP^A - G^{AB} V_{,B}/H$, we obtain the
announced exact long-wavelength result:
    \beq\label{constr2}
    \gth_i^1 = 2 \get^\perp \gz_i^2.
    \eeq
A linear version of this result can be found in \cite{vantent}, but here we
show it is valid nonlinearly as well.
It is the generalization of the well-known result \cite{conslaw} that $\gz$ is
constant in single-field inflation, showing the influence of the isocurvature
($e_2$) mode on the adiabatic ($e_1$) mode in multiple-field inflation. Note the
result is valid for an arbitrary number of fields, not just for two fields.

Hence we can now define a vector $v_{i\,a}$ without the $\gth_i^1$ component:
   \beq\label{basic_def}
    v_{i\,a}\equiv(\gz_i^1,\gz_i^2,\gth_i^2,\gz_i^3,\gth_i^3,\ldots)^T\,,
   \eeq
satisfying our key nonlinear long-wavelength evolution equation
\cite{formalism} with the constraint imposed,
    \beq\label{basic_sym}
    \dot{v}_{i\,a}(t,\vc{x}) + A_{ab}(t,\vc{x}) v_{i\,b}(t,\vc{x})
    = 0,
    \eeq
where the index $a$ runs over $2\times$(number of fields)$-1$.
For two fields on a flat manifold in the $NH=1$ gauge the matrix $A$ is given by
    \beq\label{A2f}
    A = \pmtrx{0 & -2\get^\perp & 0\\0
    &0&-1\\0&3\gc+2\ge^2+4\ge\get^\parallel+4(\get^\perp)^2+\gx^\parallel
    &3+\ge+2\get^\parallel
    }.
    \eeq
For a curved manifold with a nontrivial field metric
the term $-(2\ge/\gk^2) e_2^A R_{ABCD} e_1^B e_1^C e_2^D$ should
be added to the $A_{32}$ component, with $R^A_{\;BCD}$ the
curvature tensor of the field manifold.
Again, we stress that no slow-roll approximation has been made.

\section{Perturbative expansion}

To solve the master equation (\ref{basic_sym}) analytically, we
expand the system as an infinite hierarchy of linear perturbation
equations with known source terms at each order. To second order we
obtain
    \bea\label{vimeq1}
    \dot{v}^{(1)}_{i\,a} + A^{(0)}_{ab}(t)
    v^{(1)}_{i\,b} & = & b^{(1)}_{i\,a}(t,\vc{x}), \\
    \dot{v}^{(2)}_{i\,a} + A^{(0)}_{ab}(t)
    v^{(2)}_{i\,b} & = &
    - A^{(1)}_{ab}(t,\vc{x})v^{(1)}_{i\,b}\,,
    \label{vimeq2}
    \eea
where $v_{i\,a}^{\,}=v^{(1)}_{i\,a}+v^{(2)}_{i\,a}$, and
    \bea\label{varA}
    A_{ab}(t,\vc{x}) & = & A_{ab}^{(0)} + A_{ab}^{(1)}
    \: = \: A_{ab}^{(0)} + \der^{-2} \der^i (\der_i A_{ab})^{(1)}
    \non\\
    & \equiv & A_{ab}^{(0)}(t)
    + \bA^{(0)}_{abc}(t)v^{(1)}_{c}(t,\vc{x}).
    \eea
Here we have denoted $v^{(1)}_c \equiv \der^{-2} \der^i
v^{(1)}_{i\,c}$ and $\der_i A_{ab}$ is computed using
(\ref{constr}) and (\ref{slowrollpar}). For two fields on a flat
manifold the exact result for $\bA_{abc}$ is
\begin{widetext}
    \beq\label{Abar}
    \bA = \pmtrx{\vc{0} & \pmtrx{2\ge\get^\perp
    -4\get^\parallel\get^\perp + 2\gx^\perp\\
    -6\gc - 2\ge\get^\parallel - 2(\get^\parallel)^2 - 2(\get^\perp)^2\\
    -6 - 2\get^\parallel} & \vc{0}\\
    \vc{0} & \vc{0} & \vc{0}\\
    \vc{0} & \bar{\vc{A}}_\vc{32} & \pmtrx{-2\ge^2 - 4\ge\get^\parallel +
    2(\get^\parallel)^2 - 2(\get^\perp)^2 - 2\gx^\parallel\\
    -4\ge\get^\perp - 2\gx^\perp\\ -2\get^\perp}},
    \eeq
with
    \beq
    \bar{\vc{A}}_\vc{32} = 2\pmtrx{-6\ge\get^\parallel - 6(\get^\perp)^2
    - 3\ge\gc - 4\ge^3 - 10\ge^2 \get^\parallel
    - 2\ge(\get^\parallel)^2 - 6\ge(\get^\perp)^2
    + 8\get^\parallel(\get^\perp)^2 - 3\ge\gx^\parallel
    - 6\get^\perp\gx^\perp + \frac{3}{2} (\tV_{111}-\tV_{221})\\
    -12\ge\get^\perp - 6\get^\parallel\get^\perp + 12\get^\perp\gc
    - 6\ge^2\get^\perp + 4(\get^\parallel)^2\get^\perp
    + 4(\get^\perp)^3 - 4\ge\gx^\perp - 2\get^\parallel\gx^\perp
    + \frac{3}{2} (\tV_{211}-\tV_{222})\\
    6\get^\perp - 2\ge\get^\perp + 4\get^\parallel\get^\perp
    - 2\gx^\perp},
    \eeq
\end{widetext}
where $\tV_{lmn} \equiv (\sqrt{2\ge}/\gk) e_l^A e_m^B e_n^C V_{;ABC}/(3H^2)$.
Note that in the $NH=1$ gauge we are using here, the exact $\bA_{abc}$ only
contains terms up to next-to-leading (third) order in slow roll, no
higher-order terms exist.

The source term $b_{i\, a}^{(1)}$ on the right-hand side of the
first-order equation (\ref{vimeq1}) is not present in our original
long-wavelength equation (\ref{basic_sym}) and requires further
explanation. In this form equation (\ref{vimeq1}) is nothing more
than the full linear perturbation equation for $\gz_i^m$ rewritten
after smoothing over short wavelengths with a window function
$\cW(kR)$ with smoothing length $R \equiv c/(aH)$ and $c$ a constant 
of the order of a few.\footnote{In the gauge $t=\ln(aH)$ used in
\cite{formalism,mf}, $R$ is a function of time only (which was the
reason we chose that gauge for a stochastic numerical
implementation). In the present gauge, $t=\ln a$, which is more
convenient for super-horizon calculations, this is no longer true.
However, the spatial dependence of $H$ is of higher order in slow
roll [see (\ref{constr})], so that it can be neglected if slow
roll holds during horizon crossing.} The window function is chosen
such that: (a) $\cW(kR)\rightarrow 0$ for $k \mg 1/R$ and (b)
$\cW(kR)\rightarrow 1$ for $k\ll 1/R$. The first condition ensures
that short wavelengths are cut out, making our long-wavelength
approximations applicable, while the second ensures that at
sufficiently late times the solution of (\ref{vimeq1}) does not
depend on the exact shape of $\cW$, simply being the appropriate
linear solution on long wavelengths. The source $b_{i\,a}$ can
then be expressed in terms of the linear mode function solutions
$X^{(1)}_{am}(k,t)$ as
    \beq\label{linear_source}
    b^{(1)}_{i\,a} = \dk \, \dot{\cW}(k)
    X^{(1)}_{am}(k) \hat{a}^\dagger_m(\vc{k})
    \, \mathrm{i}k_i \,\mathrm{e}^{\mathrm{i} \vc{k}\cdot\vc{x}}
    + \mathrm{c.c.}\,.
    \eeq
Because of the conditions (a) and (b) described above, the time
derivative $\dot{\cW}(kR)$ peaks around a time just after horizon
crossing and the term $b^{(1)}_{i\,a}$ can be seen as providing
the initial conditions for the long-wavelength linear evolution of
each mode $k$. Given this behaviour of $\cW(kR)$,
the matrix $X_{am}$ in (\ref{linear_source}), containing the
Fourier solutions from linear theory for $\gz$ and $\gth$, must be
known around horizon crossing. These linear solutions can be found
directly numerically, but a general analytic solution is known in
the slow-roll approximation \cite{vantent}. In the two-field case,
the slow-roll horizon-crossing solution is simply \cite{mf}
    \beq\label{Xapprox}
    X_{am} = - \frac{\gk}{\sqrt{2} \, k^{3/2}} \frac{H}{\sqrt{2\ge}}
    \pmtrx{1&0\\0&1\\0&-\gc}.
    \eeq
The quantum creation ($\hat{a}^\dagger_m$) and conjugate
annihilation operators in (\ref{linear_source}) satisfy the usual
commutation relations.\footnote{In \cite{formalism,mf} we used a
stochastic description of the source term, necessary for
a complete numerical implementation. Here we can retain the
standard quantum description of \cite{mfb}. This leads to
different $\sqrt{2}$ factors in some intermediate steps
compared with~\cite{mf}.}

In the second-order equation (\ref{vimeq2}), the source term on
the right-hand side represents nonlinear corrections due to
long-wavelength evolution. One might be concerned that any
nonlinearity produced before and during horizon crossing
($b_{i\,a}^{(2)}$) is ignored. However, recent work
examining both tree-level \cite{seelid} (see also \cite{mf})
and quantum effects \cite{weinberg} indicates that any such nonlinearity
will be suppressed by small slow-roll factors and lead to non-Gaussianity
of the order of that produced by single-field inflation. Hence any
significantly larger non-Gaussianity will have to come from the subsequent
nonlinear evolution on superhorizon scales (which is a purely multiple-field
effect), which is why we concentrate on this regime in our treatment.

Equations (\ref{vimeq1}) and (\ref{vimeq2}) (as well as any higher
order equations), together with the initial condition
$v_{i\,a}(t\rightarrow-\infty) = 0$ implied by the formalism, can
be solved using a single Green's function $G_{ab}(t,t')$,
satisfying
    \beq\label{Green}
    \frac{\d}{\d t}\, G_{ab}(t,t') + A^{(0)}_{ac}(t) G_{cb}(t,t') = 0,
    \quad
    \lim_{t\rightarrow t'} G_{ab}(t,t') = \gd_{ab}.
    \eeq
The solution is the time integral of $G_{ab}$ contracted with the
terms on the right-hand side of (\ref{vimeq1}) and (\ref{vimeq2}).
Hence, the superhorizon linear mode functions $v_{am}^{(1)}$ are
given by
    \beq\label{vam}
    v^{(1)}_{am}(k,t) = \int_{-\infty}^t \d t' \,
    G_{ab}(t,t') \dot{\cW}(k,t') X_{bm}^{(1)}(k,t'),
    \eeq
with
    \beq
    v^{(1)}_{i\,a}(t,\vc{x}) = \int\frac{\d^3\vc{k}}{(2\pi)^{3/2}}
    \, v^{(1)}_{am}(k,t) \hat{a}_m^\dagger(\vc{k}) \, \mathrm{i} k_i
    \mathrm{e}^{\mathrm{i}\vc{k}\cdot\vc{x}} + \mathrm{c.c.}\,.
    \eeq
On superhorizon scales, the solution (\ref{vam}) is equivalent to
that from the full linear perturbation equation, so we could have
obtained it from a direct numerical calculation rather than
projecting forward our analytic solution at horizon crossing as we did
here. In particular, the integral in (\ref{vam}) is
independent of the exact form of the window function $\cW$ for
times sufficiently long after horizon crossing (in practice this
means just a few e-folds).  We note that a convenient choice for
explicit calculations is to employ a simple step function (`top
hat') as window function, so that $\dot{\cW}(k,t) =
\gd(kR/\sqrt{2}-1)$, with $R = (c/H)\exp(-t)$ the smoothing
length, which is different from the Gaussian window function used in
\cite{mf}. Finally, then, at second order we obtain the
expression
    \beq
    v^{(2)}_{i\,a}(t,\vc{x})=-\int_{-\infty}^t \d t'
    \, G_{ab}(t,t')\bar{A}_{bcd}(t')v^{(1)}_{i\,c}(t',\vc{x})
    v^{(1)}_{d}(t',\vc{x}),
    \eeq
which is again independent of the exact choice for $\cW$.

\section{Bispectrum}

So far we have used time slices for which $NH=1$, because it simplifies
superhorizon calculations, that is, time slices in which the
expansion of the universe is homogeneous ($t=\ln a$).  On such time slices
    \beq
    \label{gi_var2}
    \gz_i^1 = - \frac{\gk}{\sqrt{2\ge}} \,(e_{1A}\, \der_i \gf^A)
    = \frac{1}{3\Pi^2}\,\der_i\gr,
    \eeq
where $\gr$ is the energy density. [To derive this, one uses (\ref{constr}),
(\ref{slowrollpar}), and $H^2=\gk^2 \gr/3$.] One can show that $\gz_i^1$ is a
total gradient in the single-field case. However, for multiple fields
this is not the case and hence $\gz_i^1$ cannot be written as
the gradient of some scalar. To connect with scalar observables it
is more convenient to go to uniform energy density time slices where
$\der_i\gr=0$ \cite{LV}. On such time slices we simply have
    \beq
    \tgz_i^1=\der_i \ln \ta \equiv \der_i \tga.
    \eeq
Denoting these new time slices by $T = t+\Delta t$ and expanding
the right-hand side of (\ref{transform}) to second order we have (suppressing
the $\vc{x}$ dependence)
    \bea\label{zetagaugecorr}
    \tgz^{(1)\,1}_i(t) & = & \gz^{(1)\,1}_i(t),
    \non\\
    \tgz^{(2)\,1}_i(t) & = & \gz^{(2)\,1}_i(t)
    - \gD t \, \der_t \gz^{(1)\,1}_i.
    \eea
So no correction appears at linear order, which is as it should be, since
$\gz_i$ is gauge-invariant at linear order. At second order there is a
correction proportional to the time derivative of $\gz_i$, which is zero in the
single-field case. The quantity $\gD t$ is the time difference between uniform
expansion and uniform energy density time slices, and is evaluated by comparing
$\ga$ in the two gauges:
    \bea
    \ga(t) & = & \tga(T,\vc{x}) \: = \: \tga(t+\gD t,\vc{x})
    \non\\
    & = & \tga(t,\vc{x}) + \gD t \, \der_t \tga(t,\vc{x}) + \ldots,
    \eea
which, expanded to first order, becomes
    \beq
    \ga^{(0)}(t) = \tga^{(0)}(t) + \tga^{(1)}(t,\vc{x})
    + \gD t^{(1)}(t,\vc{x}) \der_t \tga^{(0)}(t).
    \eeq
Hence we find that $\tga^{(0)} = \ga^{(0)}$ and
    \beq
    \gD t = - \frac{\tga^{(1)}}{\der_t \ga^{(0)}} = - \gz^{(1)\,1},
    \eeq
($NH=1 \Rightarrow \der_t \ga^{(0)}=1$) so that we can rewrite
(\ref{zetagaugecorr}) as
    \beq
    \der_i \tga^{(2)} = \tgz^{(2)\,1}_i
    = \gz^{(2)\,1}_i + 2\get^\perp \gz^{(1)\,1} \gz^{(1)\,2}_i.
    \eeq

Using our formulae, we can now write down the power spectrum and
bispectrum for $\tga = \der^{-2} \der^i \tilde{\gz}_i^1$.
For the power spectrum we have
    \beq\label{powerspec}
    \left\langle \tga \tga \right\rangle^{(1)}(k,t)
    = v^{(1)}_{1m}(k,t) v^{(1)}_{1m}(k,t),
    \eeq
with $X_{bm}^{(1)}$ in (\ref{vam}) assumed to be real [as in
(\ref{Xapprox})]. By combining the different permutations of
$\langle{\tga}^{(2)}(\vc{x}_1){\tga}^{(1)}(\vc{x}_2)
{\tga}^{(1)}(\vc{x}_3)\rangle$ containing the linear and second-order
adiabatic solutions, we arrive at our main result for the connected
bispectrum:
    \bea\label{bispectrum}
    && \left\langle \tga \tga \tga \right\rangle^{(2)}
    (\vc{k}_1,\vc{k}_2,\vc{k}_3,t)\\
    && = (2\pi)^3 \gd^3({\textstyle \sum_s} \vc{k}_s)
    \Bigl[ f({k}_1,{k}_2) + f({k}_1,{k}_3)
    + f({k}_2,{k}_3) \Bigr]
    \non
    \eea
with
    \beq\label{fkk}
    f({k},{k}') \equiv
    L_{mn}(k,k',t) v^{(1)}_{1m}(k,t) v^{(1)}_{1n}(k',t)
    + {k} \leftrightarrow {k}'
    \eeq
and
    \bea\label{Lmn}
    && L_{mn}(k,k',t) \equiv \get^{\perp}(t) v^{(1)}_{2m}(k,t)
    v^{(1)}_{1n}(k',t)
    \\
    &&-\frac{1}{2}\int_{-\infty}^t \d t' \, G_{1a}(t,t')
    \bA^{(0)}_{abc}(t') v^{(1)}_{bm}(k,t') v^{(1)}_{cn}(k',t').
    \non
    \eea
This expression, valid if the non-Gaussianity produced at horizon
crossing can be neglected, gives the bispectrum in terms of
background quantities and linear perturbation quantities at horizon
crossing. It is exact: no slow-roll approximation has been used.
Since $\tgz_i^1=\der_i \tga$ is a total gradient, there are no
nonlocal terms involving inverse powers of momenta.\footnote{In
\cite{mf}, where we did not make the switch from $\gz_i^{(2)\,1}$ to
the total gradient $\tgz_i^{(2)\,1}$, the result contained factors
of the form $\vc{k}\cdot(\vc{k}+\vc{k}')/|\vc{k}+\vc{k}'|^2$. One
can easily relate that result to this one by noting that
$\vc{k}\cdot(\vc{k}+\vc{k}') = \frac{1}{2}|\vc{k}+\vc{k}'|^2 +
\frac{1}{2}(k^2-{k'}^2)$, where the second term disappears after the
correction from switching to $\partial_i\rho=0$ time slices has been
included.} In a forthcoming paper \cite{RSvTnew} we will further
work out this expression for the bispectrum analytically. Here we
finish with a numerical investigation of a specific model.

\section{Numerical investigation of a two-field quadratic potential}

We investigate a simple model with two fields and a quadratic
potential:
    \beq\label{quadratic}
    V(\phi_1,\phi_2) = {\textstyle \frac{1}{2}} m_1^2 \gf_1^2 
    + {\textstyle \frac{1}{2}} m_2^2 \gf_2^2\,,
    \eeq
choosing $m_1 = 10^{-5} \gk \inv$ (the overall mass
magnitude can be freely adjusted to fix the amplitude of the power spectrum)
and initial conditions $\gf_1 = \gf_2 = 13 \gk\inv$ for a total of about $85$
e-folds of inflation. The higher the mass ratio $m_2/m_1$ is chosen, the
sharper a turning of the trajectory in field space this model has during the
last 60 e-folds, which is reflected by a peak in $\get^\perp$, possibly
followed by additional oscillations.

    \begin{figure}
    \includegraphics[width=8cm]{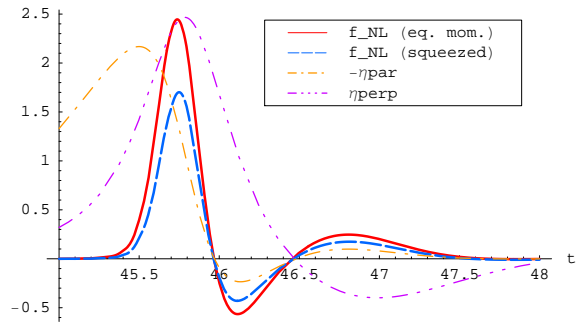}
    \caption{The non-Gaussianity parameter $\tf_\mathrm{NL}$ (\ref{fNLdef})
    for the two momentum configurations described in the text together with 
    the slow-roll
    parameters $-\get^\parallel$ and $\get^\perp$, plotted as a function of
    time in the model (\ref{quadratic}) with initial conditions
    $\gf_1=\gf_2=13\gk\inv$ and mass ratio $m_2/m_1=12$.}
    \label{fNL12fig}
    \end{figure}

In figure~\ref{fNL12fig} we show the result for the
non-Gaussianity parameter $\tf_\mathrm{NL}$, defined as \cite{mf}
    \beq\label{fNLdef}
    \tf_\mathrm{NL} \equiv
    \frac{\langle \tga \tga \tga \rangle(k_1,k_2,k_3)}
    {\left[\langle \tga \tga \rangle_{k_1} \langle \tga \tga
    \rangle_{k_2}+(k_2\leftrightarrow k_3) + (k_1\leftrightarrow k_3)
    \right]/3}
    \eeq 
[where the $(2\pi)^3$ and the $\gd$-function are removed from the
expression for the bispectrum],
in the case of a mass ratio $m_2/m_1 = 12$. It is plotted as a
function of time during the few e-folds of the corner-turning in
field space, together with $\get^\parallel$ and $\get^\perp$.
We show the results for two momentum configurations: (a) all three
momenta equal, with horizon crossing 60 e-folds before the end of
inflation, and (b) two momenta equal, crossing the horizon 60
e-folds before the end of inflation, with the third momentum
crossing the horizon 20 e-folds earlier (squeezed configuration).
We see that a large $\tf_\mathrm{NL}$ is produced initially but,
in this model, is completely erased by the time
$\get^\perp$ has its first zero.\footnote{Because of subtle
cancellations between terms of different order in slow roll, the
leading-order slow-roll estimate in \cite{mf} (for the case of
mass ratio 9) overestimates the non-Gaussianity substantially,
although it does provide a fair estimate for the maximum value
reached during inflation.} The dependence on different
momentum configurations can be inferred from (\ref{Xapprox}) and
(\ref{vam}). The main momentum dependence is the overall factor of 
$k^{-3/2}$ of $v_{am}^{(1)}$; however, if other momentum dependence were
absent, these factors would cancel exactly in the definition of 
$\tf_\mathrm{NL}$.
The remaining dependence on scale comes from the slow-roll suppressed 
dependence of $X_{am}$ on time and the fact that $\dot{\cW}(k,t)$ 
peaks at different times for different modes. Because of this slow-roll
suppression the momentum dependence is rather weak; figure~\ref{fNL12fig}
shows a variation of less than a factor 2 between two very different
configurations.

The large non-Gaussianity of the type described above might be
accessible in models where inflation ends during the
corner-turning, such as hybrid inflation, or in which residual
isocurvature modes persist. We have also investigated quartic
potentials, and find that large, but temporary, non-Gaussianity
emerges in a similar fashion.  It may be that persistent large
non-Gaussianity is easier to achieve in less symmetric potentials.
Analytic calculations in \cite{RSvTnew}, as well as further
numerical investigations that are currently underway, should
settle these issues.

\section{Discussion}

In this paper we extended and improved the non-Gaussian formalism developed
in our earlier work. We derived and made use of a long-wavelength constraint 
relation (\ref{constr2}), which simplifies the system of equations 
derived in \cite{mf,formalism}. We also clarified the issue of different 
time slicings at second order in order to properly connect to scalar 
observables. 
Using this formalism we have provided a general, exact expression for the
bispectrum in multifield inflation due to superhorizon effects, equations
(\ref{bispectrum})--(\ref{Lmn}), which provides a powerful way to compute
non-Gaussianity for an arbitrary inflationary model up to the end
of inflation. The computation is technically straightforward since
it only requires the background solution and the linear perturbation
solution at horizon crossing to be known, either analytically or
numerically.

A numerical investigation of the simplest two-field
quadratic example showed that non-Gaussianity can become large, as 
parametrized by a value of $\tf_\mathrm{NL}$ of order unity, whereas 
single-field inflation generically predicts a result two orders of magnitude 
smaller \cite{maldacena}. In this model a temporary breakdown of slow roll 
is required somewhere during the last 40--50 e-folds of inflation (i.e.\ some 
time after the observable scales have left the horizon so that the scalar 
spectral index will not be too far from unity). This happens for mass ratios 
larger than about $10$ when the trajectory of the scalar fields driving
inflation turns a corner, as parametrized by $\get^\perp$.
The large non-Gaussianity is produced by the nonlinear influence
of the isocurvature mode on the adiabatic mode but it also decays away as
the isocurvature mode disappears.
The exact behaviour of $\tf_\mathrm{NL}$ as a function of time during inflation
is complicated by subtle cancellations and a full explanation is left for a
future publication \cite{RSvTnew}.
We find that there is only a weak dependence on the specific momentum
configuration.

We note that earlier work in \cite{beruza} emphasised the
importance of the relation between non-Gaussianity and isocurvature
modes, although their models relied
on nonlinear interacting potentials to generate higher-order correlators.
Related work in
\cite{wandsvernizzi} with the `$\delta N$' formalism, using the
same quadratic two-field potential, produced results which are
qualitatively similar, that is, with large non-Gaussianity emerging
temporarily. Numerically the $\delta N$ approach evolves
neighbouring inflationary trajectories through a multidimensional field space,
with appropriate derivatives estimated from these. In contrast, our approach 
relies more directly on integrals over the linear perturbation and background 
solutions which, in principle, should be more accurate.
A direct quantitative comparison and an analysis of the
relative merits of the two approaches deserves further
investigation.

Attempts to incorporate inflation within the framework of a more
fundamental physical theory like string theory or extensions of the
Standard Model invariably lead to inflationary models with many
scalar fields. Our findings indicate that non-Gaussianity can emerge
naturally in these models, although its perdurance seems closely
linked to that of the isocurvature modes. It is of great interest to
investigate scenarios in which this non-Gaussianity might survive to
late times because its imprint in the CMB may be observable in
forthcoming experiments.

\section{Acknowledgements}

We are grateful for fruitful discussions with Neil Barnaby, Francis
Bernardeau, James Cline, Jean-Philippe Uzan, Filippo Vernizzi, and
David Wands. This research was supported by PPARC Grant No.\ PP/C501676/1.

\end{document}